\documentclass[12pt]{article}
\usepackage{graphicx}
\usepackage{epsfig}
\begin{document}
\title{Stabilizing near-nonhyperbolic chaotic systems and its potential applications in neuroscience}
\author{{
Debin Huang \footnote{E-mail: dbhuang@mail.shu.edu.cn}
}\\\\
{\small Department of Mathematics, Shanghai University, Shanghai
200436, P.R. China}}
\date{}
\maketitle
\begin{center}
\begin{minipage}{12cm}
\hskip 1cm Based on the invariance principle of differential
equations a simple, systematic, and rigorous feedback scheme with
the variable feedback strength is proposed to stabilize
nonlinearly any chaotic systems without any prior analytical
knowledge of the systems. Especially the method may be used to
control near-nonhyperbolic chaotic systems, which although arising
naturally from models in astrophysics to those for neurobiology,
all OGY-type methods will fail to stabilize. The technique is
successfully used to the famous Hindmarsh-Rose model neuron and
the R$\ddot{\textrm{o}}$ssler
hyperchaos system.\\
\vskip 0.01cm PACS number(s): 05.45.Gg; 87.10.+e; 87.19.La
\end{minipage}
\end{center}
Since Ott, Grebogi, and York (OGY) [1] proposed an effective
method of chaos control, all kinds of variations based on this
method have been given [2], and lots of successful experiments
have been reported. For the simplicity, here we call the OGY
method and its variations as the OGY-type methods. Recall the idea
of the OGY-type methods, the following three steps are necessary
to improve its performance: (i) To specify and locate an unstable
periodic orbit embedded
 in the chaotic attractor, say a fixed point
 $x_f$; (ii) To approximate linearly the system in a small neighborhood of
 $x_f$ by reconstructing statistically the corresponding linearized matrix $J$; (iii) To control (or
stabilize) the chaotic orbits entering the neighborhood to $x_f$
 with aid of the approximated linear dynamics. The first step can be realized by the method
of close returns from experimental data, and the third step is
entirely within the field of linear control theory such as ``the
pole placement technique". Although the second step, including the
calculation of eigenvalues and eigenvectors of the corresponding
linearized matrix $J$, has been solved by the least-squares fit,
this problem is related to how chaos affects the linear estimation
of the dynamics in the small neighborhood of
 $x_f$. Especially, when the system is nonherpybolic and any prior analytical
knowledge of dynamics is not available, such linearization will be
problematic due to the nonlinearity. It is well known that many of
the chaotic phenomena seen in systems occurring in practice are
nonhyperbolic. On the other hand, there are numerous successful
reports of OGY control in numerical experiments. This matter is
slightly puzzling. In [3], the author investigated carefully this
problem, and found that there are two possible reasons resulting
in such contradiction. One reason is that the least-squares fit
used in the process of reconstructing the attractor from time
series is ill-defined due to the nonhyperbolicity of system. The
other is that there are large relative errors in the process of
solving numerically eigenvalues of a matrix as one of its
eigenvalues, $\lambda\approx 0$, which is a well-known fact in the
matrix computations [4]. Therefore in those successful numerical
experiments the nonhyperbolicity of system may be destroyed before
obtaining the information for attempting the OGY control by
experimental time series. Although the nonhyperbolicity of the
chaotic attractor does not automatically mean the nonhyperbolicity
of the unstable periodic orbits embedded in it, the
near-nonhyperbolicity must exist and affect the performance of the
OGY control when the system itself is near-nonhyperbolic, see the
models discussed below. More interestingly, the report in [5], on
failure of chaos control in a parametrically excited pendulum
whose excitation mechanism is not perfect, throws highly the light
to this viewpoint (to the best of my knowledge, this is the first
report on failure of the OGY control in the concrete physical
experiments).
\par This letter is motivated by the
limitation of the OGY-type controllers as what referred above.
Especially we address the control problem on the
near-nonhyperbolic chaotic systems in the form of
$$\dot u=g_u(u,v),\ \ \ \ \ \dot v=r g_v(u,v),
\eqno (1)$$ where $u\in \textrm{R}^{n_1}$, $v\in
\textrm{R}^{n_2}$, and $0<r\ll 1$. The systems have simultaneously
$n_1$ dramatic components $u$ and $n_2$ slow variables $v$, which
arise naturally from many scientific disciplines, and range from
models in astrophysics to those for biological cells [6]. In
particular, such systems and their discrete versions are widely
used to model bursting, spiking, and chaotic phenomena in
neuroscience, see [7] and references therein. More interestingly,
just as what referred in [1], due to multipurpose flexibility of
higher life form, chaos may be a necessary ingredient in their
regulation by the brain. In [3] the author guessed that such
chaotic ingredient is probably in the form of (1), where the slow
variables represent a ``container" or``recorder" storing the
acquired knowledge, which is attached on some neurons. The guess
is mainly based on a fact of cognitive science, namely, the
acquisition of some knowledge will give way to acquire other
knowledge in the brain, and hence the acquisition of knowledge
will ``decrease" the freedom of topology structure of the brain
[8]. If the chaotic ingredient in the brain is from the systems in
the form of (1), then all OGY-type methods will fail to control
such chaotic dynamics because the systems with sufficiently small
$r$ are near-nonhyperbolic (actually note that in discrete case if
the linearization matrix $J$ admits one eigenvalue
$\lambda\approx1$ (resp. $\lambda\approx0$ for the continuous
case) all OGY-type controllers are infeasible because those
controllers contain the near-singular term $(J-I)^{-1}$, see [2]).
This is just the reason for failure of chaos control reported in
[5]. In the meantime, this reflects the fact that the slow varies,
i.e., the neurons recording the knowledge, should not be ignored
for achieving multipurpose flexibility of the brain because the
related knowledge has to be excited to respond to the signals
entering the brain. Moreover, this mechanism is beneficial to
explain why stabilization of an inverted triple pendulum is very
troublesome as out-of planar motions become very substantial,
which was firstly reported in [9].  In this letter, based on the
invariance principle of differential equations [10], a simple and
rigorous feedback scheme with the variable feedback strength is
proposed to stabilize nonlinearly any chaotic and hyperchaotic
systems without any prior analytical knowledge of the systems.
Especially, this simple technique can be easily applied to
stabilize near-nonhyperbolic chaotic systems in the form of (1).
This letter is mainly focused on the continuous systems, but the
proposed method can be generalized to the case of the discrete
version by the invariance principle of difference equations.\par
Let a chaotic system be given as
$$\dot x=f(x), \eqno(2)$$
where $x=(x_1,x_2,\cdots,x_n)\in \textrm{R}^n$, $f(x)=(f_1(x),
f_2(x),\cdots,f_n(x)):\textrm{R}^n\rightarrow \textrm{R}^n$ is a
nonlinear vector function. Without loss of the generality we let
$\Omega\subset\textrm{R}^n$ be a chaotic bounded set of (1) which
is globally attractive, and suppose that $x=0$ is a fixed point
embedded in $\Omega$. For the vector function $f(x)$, we give a
general assumption. \par {\sl For any $x=(x_1,x_2,\cdots, x_n)\in
\Omega$,
 there exists a
constant $l>0$ satisfying $$\mid f_i(x)\mid\leq l
\textrm{max}_j\mid x_j\mid, i=1,2,\cdots, n.\eqno (3)$$} \par
Note
this condition is very loose, for example, the condition (3) holds
as long as ${\partial f_i\over
\partial x_j} (i,j=1,2,\cdots, n)$ are bounded. Therefore the class of systems in the form of (2)-(3)
include almost all well-known chaotic and hyperchaotic systems. To
stabilize the chaotic orbits in (2) to the fixed point $x=0$, we
consider the feedback control
$$\dot x=f(x)+\epsilon x. \eqno(4)$$
Instead of the usual linear feedback, the feedback strength
$\epsilon=(\epsilon_1,\epsilon_2,\cdots, \epsilon_n)$ here will be
duly adapted according to the following update law:
$$
\dot {\epsilon}_i=-\gamma_ix_i^2,i=1,2,\cdots,n,\eqno(5)
$$
where $\gamma_i>0, i=1,2,\cdots,n,$ are arbitrary constants. For
the system consisting of (4) and (5), we introduce the following
function
$$
V={1\over 2}\sum_{i=1}^nx_i^2+{1\over
2}\sum_{i=1}^n{1\over\gamma_i}(\epsilon_i+L)^2, \eqno(6)
$$
where $L$ is a constant bigger than $nl$, i.e., $L>nl$. By
differentiating the function $V$ along the trajectories of the
system (4)-(5), we obtain
$$
\dot V
=\sum_{i=1}^nx_i(f_i(x)+\epsilon_ix_i)-\sum_{i=1}^n(\epsilon_i+L)x_i^2
\leq(nl-L)\sum_{i=1}^nx_i^2\leq 0. \eqno (7)
$$
where we have assumed $x\in \Omega$ (without loss of the
generality as $\Omega$ is globally attractive), and used the
condition (3). It is obvious that $\dot V=0$ if and only if $x=0
$, namely the set $E=\{(x,\epsilon)\in
\textrm{R}^{2n}:x=0,\epsilon=\epsilon_0\in \textrm{R}^{n}\}$ is
the largest invariant set contained in $\dot V=0$ for the system
(4)-(5). Then according to the well-known invariance principle of
differential equations [10], starting with arbitrary initial
values of the system (4)-(5), the orbit converges asymptotically
to the set $E$, i.e., $x\rightarrow 0$ and $\epsilon\rightarrow
\epsilon_0$ as $t\rightarrow\infty$. \par Namely, when the chaotic
system (2) is stabilized to $x=0$ the variable feedback strength
 $\epsilon$ will be automatically adapted to a suitable strength $\epsilon_0$ depending on the initial
 values. This is significantly different from the usual linear
 feedback, and the converged strength must be of the
 lower order than those used in the constant gain schemes. But theoretically
 the converged strength may be very big so that it may give rise to its own dynamics.
However the
 flexibility of the strength in the present scheme can overcome
 this limitation once such case arises. For example, suppose that the
 feedback strength is restricted not to exceed a critical value,
 say $k$. In the present control procedure, once the variable strength $\epsilon$ exceeds $k$ at
 time $t=t_0$, we may choose the values of variables at this time
 as initial values and repeat the same control by resetting the
 initial strength $\epsilon(0)=0$. Namely one may achieve the
 stabilization within the restricted feedback strength due to
 the global stability of the present scheme. This idea is slightly similar to
 that of the OGY control [1], i.e., small parameter control, but there exists a certain difference,
 for example, in the OGY control the controller waits passively for the emergence of chaotic
 orbits. In the other side, in the present scheme the small converged
 strength may be obtained by adjusting suitably the parameter
 $\gamma$. Moreover, we note that in the present scheme it is not necessary for some particular
models to use all the
 variables of the system as feedback signals. For example, one may set
$\epsilon_i\equiv0$ if $\mid e_i\mid\leq\mid e_j\mid$, and this
case exists in general due to the nonhyperbolicity of chaotic
attractor, see the following examples.
 Obviously this simple, systematic, and rigorous
 method may  stabilize nonlinearly almost all chaotic
 systems without any priori analytical knowledge of systems, and
 is robust against the effect of noise due to the global nonlinear
 stability.
\par
 Next we will give two illustrative examples. We think that the chaotic ingredient in the brain is from the systems in
the form of (1), and thus we take the famous Hindmarsh-Rose model
neuron [11] as the first example, which is governed by the
following three-order ordinary differential equation
$$\dot{x}_1=x_2+3x_1^2-x_1^3-x_3+I, \ \ \ \dot{x}_2=1-5x_1^2-x_2,\ \ \
 \dot{x}_3=-rx_3+4r(x_1+1.6)
 \eqno(8)$$
with $0<r\ll 1$. Here $x_1$ is the membrane potential of the
neuron, $x_2$ is a
 recovery variable and $x_3$ is a slow adaptation current. It has
 been found in [12] that the model admits a chaotic attractor with
 $r=0.0012$ and the external current $I=3.281$, see Figure 1 for the chaotic time series of $x_1$. After
 transforming only one fixed point $(-0.6835,-1.3359,3.666)$ to
 $(0,0,0)$, we stabilize successfully this near-nonhyperbolic chaotic system
 by the proposed scheme, where let $\epsilon_2\equiv 0$. The
 corresponding numerical results and the evolution of $\epsilon$
 are shown in Figure 2, where the initial values are set as
 $(-0.5,-0.3,0.1,0,0)$ with $\gamma_1=0.01,\gamma_3=0.1$. In
 addition, in Figure 3 we show that this chaotic system may also
 be stabilized by another two feedback signals $x_2$ and $x_3$, where
 all initial values are same those in Figure 2 and parametrical values are set as $\gamma_2=0.01,\gamma_3=0.1$.
 However we find numerically that such stabilization is troublesome by
 the feedback signals $x_1$ and $x_2$, which confirms our
 viewpoint referred above, namely, it is difficult to stabilize the near-nonhyperbolic chaotic systems
  if this near-nonhyperbolicity (resp. the slow variable) is
 ignored.
  \par
 To show the generality of the present method, our second example
 is the famous
R$\ddot{\textrm o}$ssler hyperchaos system:
$$
  \dot {x}_1=-x_2-x_3,\ \ \dot {x}_2=x_1+0.25x_2+x_4,
\ \ \dot {x}_3=3+x_1x_3,\ \  \dot {x}_4=-0.5x_3+0.05x_4.
 \eqno(9)$$
 Accordingly after transforming only one fixed point
 $(-5.4083,-0.5547,0.5547,5.547)$ to $(0,0,0,0)$, the hyperbolic
 chaotic system is stabilized by the present method, where let $\epsilon_1=\epsilon_3\equiv
 0$. The
 corresponding numerical results and the evolution of $\epsilon$
 are shown in Figure 4, where the initial values are set as
 $(5,30,5,10,0,0)$ with $\gamma_2=\gamma_4=0.2$. \par
In conclusion, we have given a simple, systematic, and rigorous
method to stabilize nonlinearly any chaotic systems. Especially
the method may be used to near-nonhyperbolic chaotic system, which
all OGY-type methods will fail to stabilize. Perhaps this idea may
explain more reasonably the multipurpose flexibility of higher
life form due to chaotic ingredient of the brain. In addition,
this idea of control has been applied successfully to chaotic
synchronization by the author [13], so we believe that the idea
may be used to explore the interesting dynamical properties found
in neurobiological systems, i.e., the onset of regular bursts in a
group of irregularly bursting neurons with different individual
properties [14].\par
 {\bf Acknowledgments:} This work is
supported by the National Natural Science Foundation (10201020).
 
\newpage

\begin{figure}
\begin{center}\epsfig{file=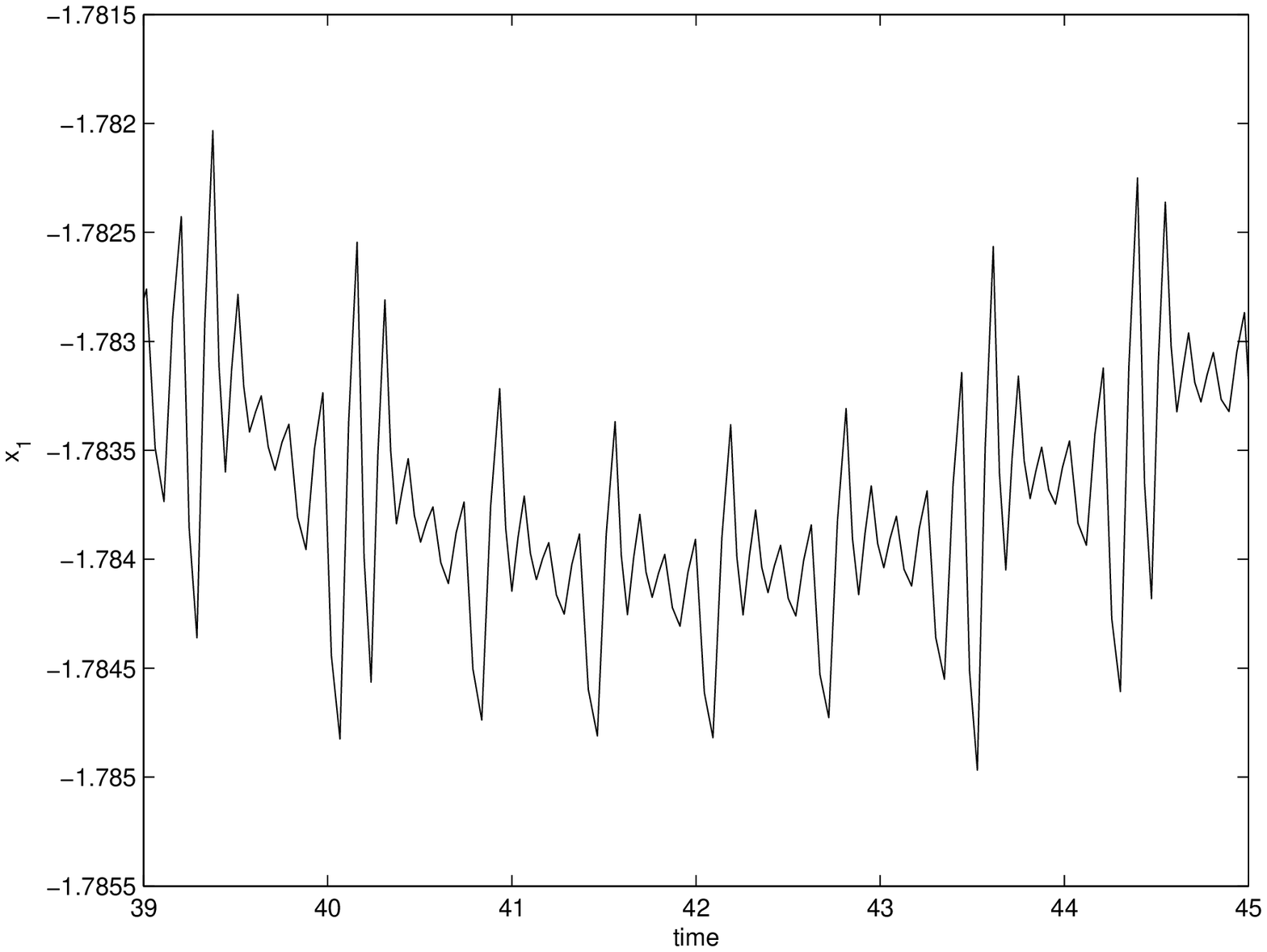,height=8cm,width=8cm}\end{center}
 \begin{center}\begin{minipage}{8cm}{\footnotesize {\bf FIG.1}.
 Time series $x_1(t)$ generated by the chaotic Hindmarsh-Rose model (8).}\end{minipage}\end{center}
\end{figure}

\begin{figure}
\begin{center}\epsfig{file=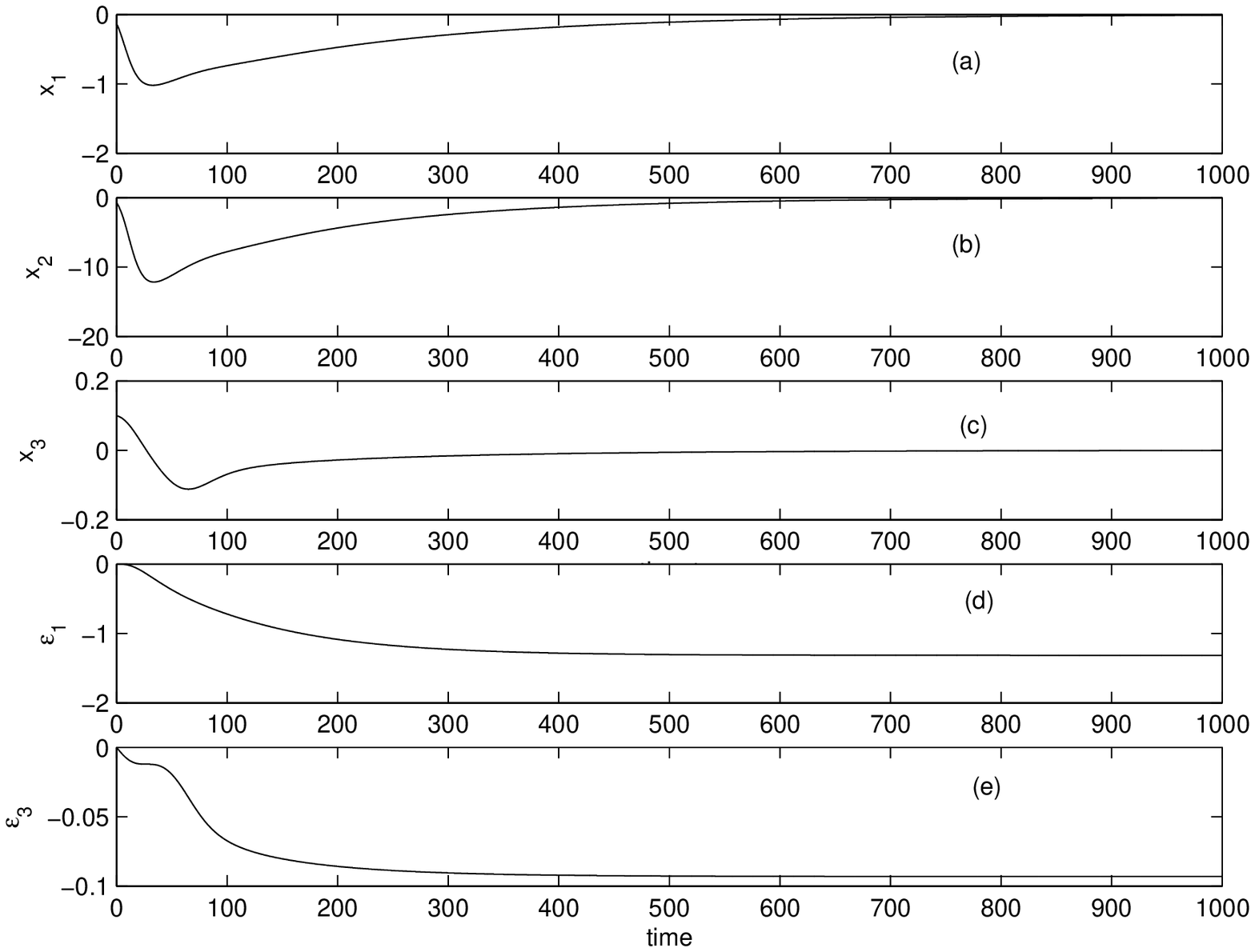,height=8cm,width=8cm}\end{center}
 \begin{center}\begin{minipage}{8cm}{\footnotesize {\bf FIG.2}.
 The chaotic Hindmarsh-Rose model (8) is stabilized successfully by only two feedback signals $x_1$ and $x_3$, where (a)-(c)
 show the temporal evolution of the variables $x_i,i=1,2,3$, and (d)-(e) correspond to the variable feedback strength $\epsilon_1$ and $\epsilon_3$.}\end{minipage}\end{center}
\end{figure}

\begin{figure}
\begin{center}\epsfig{file=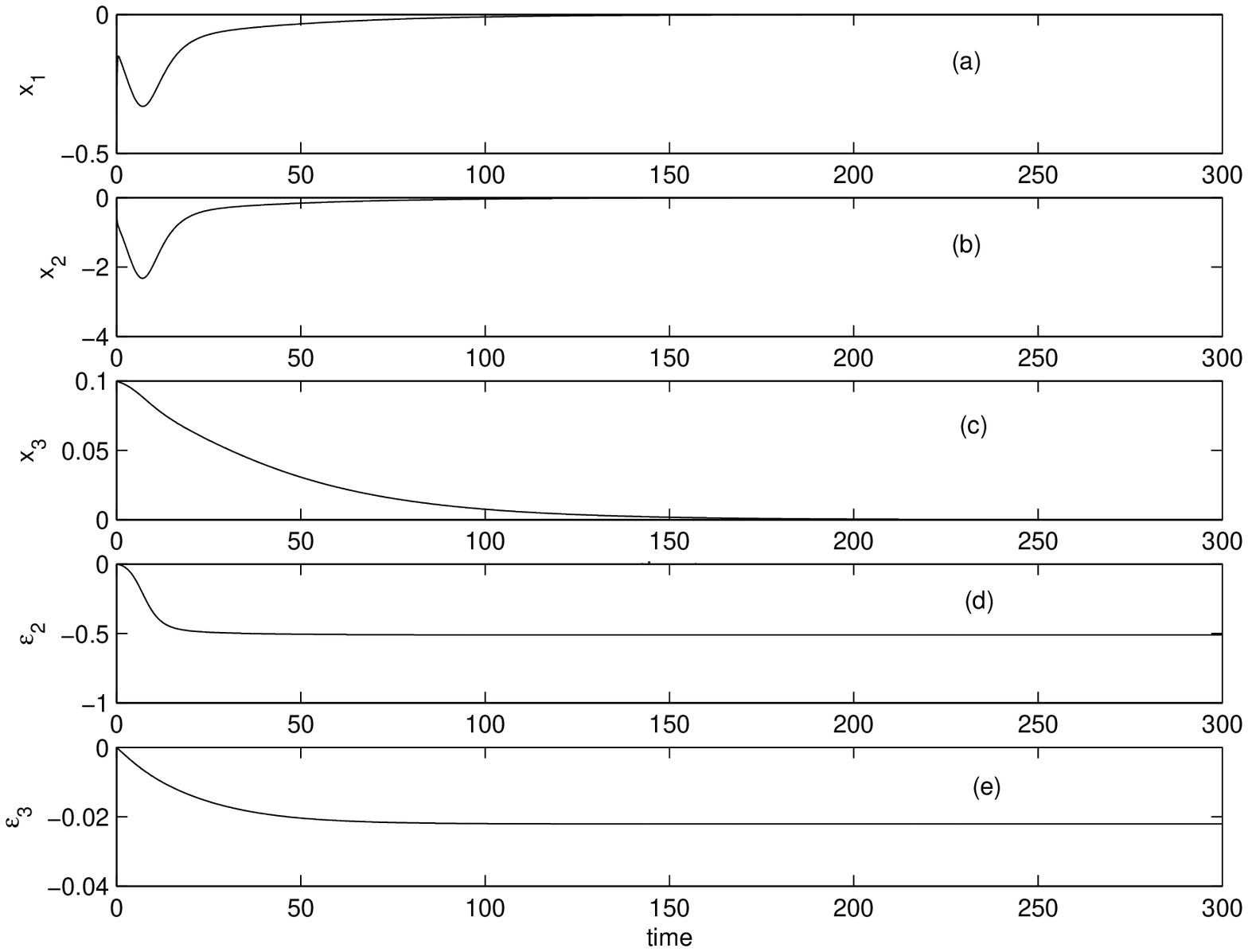,height=8cm,width=8cm}\end{center}
 \begin{center}\begin{minipage}{8cm}{\footnotesize {\bf FIG.3}.
 The chaotic Hindmarsh-Rose model (8) may also be stabilized by another two feedback signals $x_2$ and $x_3$.}\end{minipage}\end{center}
\end{figure}

\begin{figure}
\begin{center}\epsfig{file=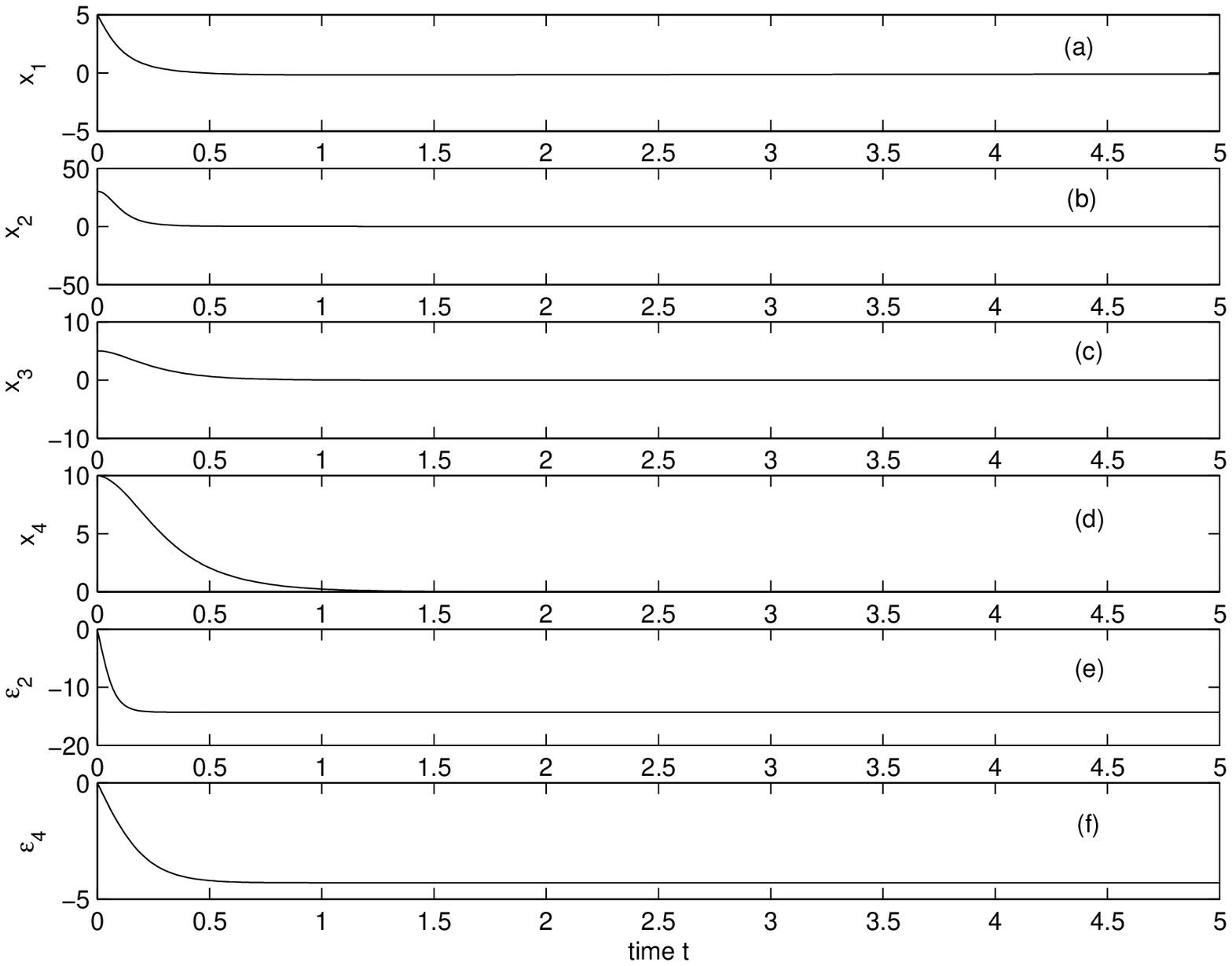,height=8cm,width=8cm}\end{center}
 \begin{center}\begin{minipage}{8cm}{\footnotesize {\bf FIG.4}.
The R$\ddot{\textrm o}$ssler hyperchaotic system (9) is stabilized
by only two feedback signals $x_2$ and $x_4$, where (a)-(d)
 show the temporal evolution of the variables $x_i,i=1,2,3,4$, and (e)-(f) correspond to the temporal evolution of feedback strength $\epsilon_2$ and $\epsilon_4$.}\end{minipage}\end{center}
\end{figure}

\begin{thebibliography}{14.8cm}
\bibitem {h04-} In other words, the neuron attaching the
knowledge is little active in comparision with other neurons
except when the related signals enter the brain, thus it will not
die out and its dynamics varies slowly. Linking with the models of
neurobiology, such guess should be reasonable.
\end{thebibliography}
\end{document}